\documentclass[conference]{IEEEtran}
\newlength{\flexwidth}
\usepackage{silence}
\usepackage[table]{xcolor}
\usepackage{cite} 
\usepackage{amsmath,amssymb,amsthm,fixmath}
\usepackage{mathtools}
\usepackage{accents}
\usepackage{xcolor}
\usepackage{siunitx}
\usepackage{cuted}
\usepackage{multirow,booktabs}
\usepackage{wrapfig}
\usepackage{subcaption}
\usepackage[inline]{enumitem}
\usepackage{optidef}
\usepackage{graphicx}
\usepackage{epstopdf}
\usepackage{lipsum}
\usepackage{amssymb}
\usepackage{diagbox}
\usepackage{adjustbox}
\usepackage{tcolorbox}
\usepackage{float}

\usepackage[normalem]{ulem}

\usepackage{dblfloatfix}
\pdfoutput=1
\usepackage[linesnumbered,ruled,vlined]{algorithm2e}

\usepackage{graphicx}

\usepackage{epstopdf}
\usepackage[textsize=tiny,colorinlistoftodos]{todonotes}
\makeatletter
\define@key{todonotes}{bh}[]{
	\setkeys{todonotes}{author=\textbf{Bin - notes}, color=red!30}}%
\define@key{todonotes}{bh2}[]{
		\setkeys{todonotes}{author=\textbf{Bin - urges}, color=green!30}}%
\define@key{todonotes}{zf}[]{
	\setkeys{todonotes}{author=\textbf{Zexin}, color=blue!30}}%
\makeatother

\usepackage{algpseudocode}

\usepackage[shortcuts,acronym,automake]{glossaries}
\makeglossaries
\newacronym{ue}{UE}{User Equipment}
\newacronym{bs}{BS}{base station}
\newacronym{csi}{CSI}{Channel state information}
\newacronym{b5g}{B5G}{Beyond-Fifth-Generation}
\newacronym{6g}{6G}{Sixth Generation}
\newacronym{ml}{ML}{Machine learning}
\newacronym{sbs}{SBS}{small base station}
\newacronym{mu}{MU}{mobile user}
\newacronym{mbs}{MBS}{macro base station}
\newacronym{mse}{MSE}{Mean Squared Error}
\newacronym{cl}{CL}{centralized learning}
\newacronym{uav}{UAV}{unmanned aerial vehicle}
\newacronym{bme}{BME}{Bayesian Model Ensemble}
\newacronym{iid}{IID}{independent and identically distributed}
\newacronym{raf}{RAF}{robust aggregation function}
\newacronym{sgd}{SGD}{stochastic gradient descend}
\newacronym{cdf}{CDF}{cumulative distribution function}
\newacronym{lid}{LID}{local intrinsic dimensionality}
\newacronym{llpf}{LLPF}{local loss pre-filtering}
\newacronym{mitm}{MITM}{man-in-the-middle}
\newacronym{ae}{AE}{adversary entitie}
\newacronym{tof}{TOF}{time of fly}
\newacronym{rssi}{RSS}{received signal strength}
\newacronym{3d}{3D}{three dimensional}
\newacronym{aoa}{DoA}{Direction of Arrival}
\newacronym{sdp}{SDP}{semi-definite programming}
\newacronym{nlos}{NLOS}{Non-Line-of-Sight}
\newacronym{snr}{SNR}{Signal to Noise Ratio}
\newacronym{crb}{CRB}{Cramer-Rao bound}
\newacronym{lse}{LSE}{least squared estimation}
\newacronym{wlse}{WLSE}{weighted least squared estimation}
\newacronym{gd}{GD}{Gradient descend}
\newacronym{ap}{AP}{Access Points}
\newacronym{crlb}{CRLB}{Cramér-Rao Lower Bound}
\newacronym{tdoa}{TDoA}{Time Difference of Arrival}
\newacronym{sinr}{SINR}{Signal to Interference and Noise Ratio}
\newacronym{los}{LOS}{Line of Sight}
\newacronym{a2g}{A2G}{Air to Ground}
\newacronym{eu}{EU}{European Union}
\newacronym{umiav}{UMi-AV}{Urban Micro–Aerial Vehicle}
\newacronym{3gpp}{3GPP}{3rd Generation Partnership Project}
\newacronym{lae}{LAE}{Low Altitude Economy}
\newacronym{gnss}{GNSS}{Global Navigation Satellite System}
\newacronym{rf}{RF}{Radio Frequency}
\newacronym{gpdr}{GDPR}{General Data Protection Regulation}

\DeclareSIUnit{\belmilliwatt}{Bm}
\DeclareSIUnit{\dBm}{\deci\belmilliwatt}

\usepackage{tcolorbox}
\tcbuselibrary{many}

\usepackage[free-standing-units=true]{siunitx}

\begin{document}
	
	\title{Lightweight Node Selection in Hexagonal Grid Topology for TDoA-Based UAV Localization}
	
	\author{\IEEEauthorblockN{Zexin~Fang\IEEEauthorrefmark{1},~Bin~Han\IEEEauthorrefmark{1}, ~Wenwen~Chen\IEEEauthorrefmark{1},~and~Hans~D.~Schotten\IEEEauthorrefmark{1}\IEEEauthorrefmark{2}}
		\IEEEauthorblockA{
  \IEEEauthorrefmark{1}{University of Kaiserslautern-Landau (RPTU), Germany}\\\IEEEauthorrefmark{2}{German Research Center for Artificial Intelligence (DFKI), Germany}
		}
	}
	
	\bstctlcite{IEEEexample:BSTcontrol}
	
	\maketitle

	\begin{abstract} This paper investigates the optimization problem for \gls{tdoa}-based \gls{uav} localization in low-altitude urban environments with hexagonal grid node deployment. We derive a lightweight optimized node selection strategy based on only \gls{rssi} measurements, to pre-select optimal nodes, avoiding extensive \gls{tdoa} measurements in energy-constrained \gls{uav} scenarios. Theoretical and simulation results demonstrate that dynamically selecting the number of reference nodes improves localization performance while minimizing resource overhead.
		
	\end{abstract}
	
	
	\IEEEpeerreviewmaketitle
	
	\glsresetall

	\section{Introduction}\label{sec:introduction}
    The \gls{lae} represents an emerging economic sector that leverages low-altitude airspace for commercial and social aviation operations. It includes diverse types of aircraft conducting missions in areas such as transportation, logistics, tourism, agriculture, and disaster monitoring \cite{LAE2025fang}. As low-altitude airspace becomes increasingly utilized for diverse flight activities, accurate localization, particularly for \glspl{uav}, grows critical for effective task execution and collision avoidance, especially when \gls{gnss} proves unreliable. 

    Current high-accuracy \gls{uav} localization solutions include: \begin{enumerate*}[label=\emph{\roman*)}]
    \item computer-vision based solution;
    \item sensing based solution;
    \item \gls{rf} based self-localization.
    \end{enumerate*} Computer-vision based solutions offer flexibility \cite{disadv2022khe} but face daytime and legal constraints, such as \gls{gpdr} in Germany. Sensing-based solutions provide high-accuracy localization by sharing position information, removing computational burden from the aircraft. However, both approaches introduce latency due to heavy computational requirements and are highly distance-sensitive, therefore limiting applications for high-mobility \glspl{uav}.
    Compared to these solutions, \gls{rf} based self-localization is lightweight and universally applicable. It captures parameters from multiple references, such as \gls{rssi}, \gls{tdoa}, or \gls{aoa}, and combines these measurements with reference positions for self-localization. \gls{tdoa}-based methods are commonly preferred due to their simpler implementation compared to \gls{aoa}-based methods and higher accuracy than \gls{rssi}-based methods \cite{tdoa2022sinha}.
    
    Most existing research assumes dedicated sensor networks with synchronized nodes to support accurate localization. Recent work has addressed energy-efficient node selection  \cite{sensor2019zhao, sensor2022zhao,sensor2021dai}, often focusing on classifying nodes by \gls{nlos} and \gls{los} conditions. However, less attention has been given to the deployment topology of these sensors in practical scenarios. While the \gls{crb} indicates that localization accuracy improves with more sensor nodes, this is constrained by sensor deployment in practice. Without comprehensive link adaptation, including distant nodes may degrade accuracy due to unreliable \gls{tdoa} measurements.
    Therefore, inspired by previous research, we investigate the impact of node deployment on localization accuracy. We consider a hexagonal grid topology for reference node placement based on two considerations. First, hexagonal grids provide uniform coverage for practical wireless deployments. Second, with \gls{6g} networks, base stations are becoming smaller and more densely deployed, making high-accuracy \gls{tdoa}-based localization increasingly feasible. While fewer works focus on base station-based \gls{tdoa} localization, with synchronization being the primary challenge \cite{baseTdoa2015sang, baseTdoa2022fan, baseTdoa2024Motie}, embedding localization nodes within base station infrastructure is compelling as it requires no additional deployment.
    
    In Sec.\ref{sec:intro}, we investigate the \gls{a2g} model to provide a foundation for subsequent optimization analysis. In Sec.\ref{sec:sys_model}, we introduce the \gls{crb} of \gls{tdoa} measurement, formulate the objective function for self-localization optimization, and analyze this optimization problem theoretically. In Sec.\ref{sec:method}, we introduce an algorithm to find optimal node selection using only \gls{rssi}, avoiding unnecessary \gls{tdoa} measurements to reduce localization latency. We validate our optimization framework and algorithm in Sec.\ref{sec:simu} and conclude in Sec.~\ref{sec:conclu}.

    \section{channel modeling}\label{sec:intro}
\begin{figure}[!ht]
    \centering
        \centering
        \includegraphics[width=0.73\linewidth]{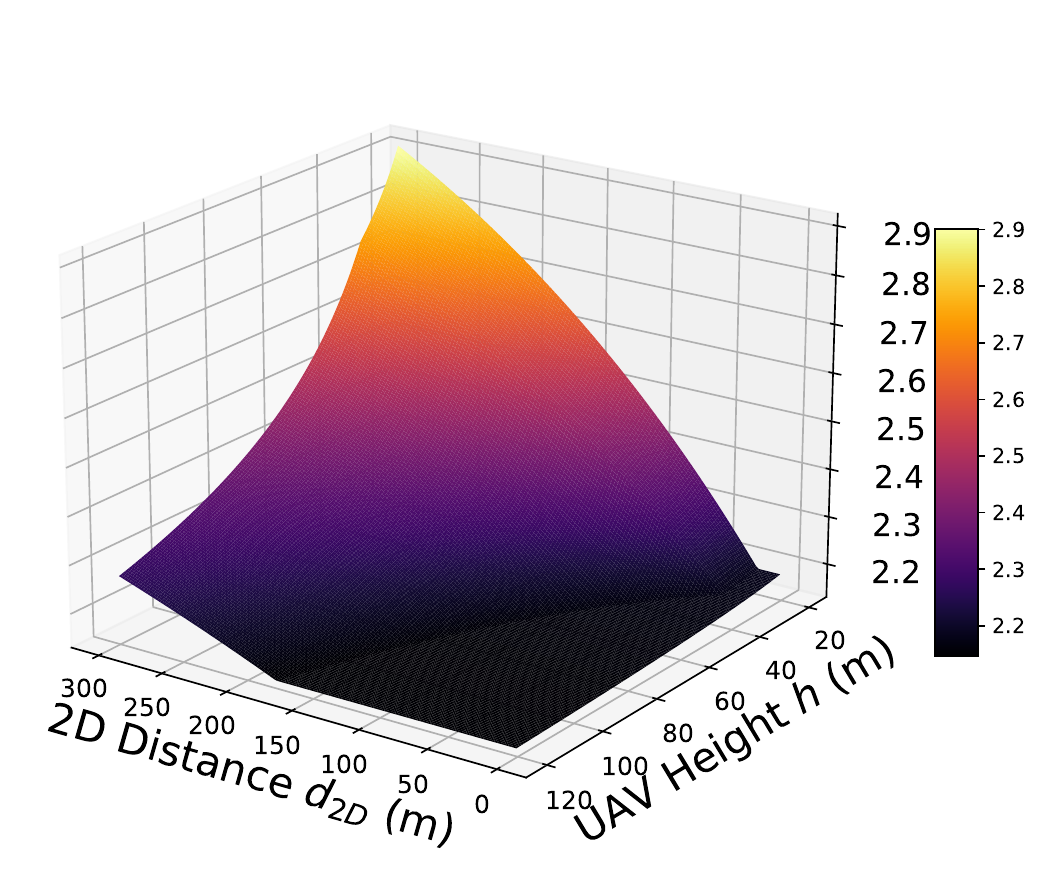}
    \caption{Path loss component $\eta$ with respect to $d_{\text{2D}}$ and $h$ (A1 is the black plane and A2 is the curved surface)}
    \label{fig:a2gchannel}
\end{figure}

    The \gls{a2g} channel characteristics were extensively investigated by \gls{3gpp} in \cite{3gpp-tr36.777}. In the \gls{umiav} channel model, the probability of \gls{los} is described with:
     \begin{equation}\label{eq:problos}
        P_\text{los}=  \begin{cases}
       1 ,\quad d_{\text{2D}} \leq d_1 ;\\
       \left(1-\frac{d_1}{d_{\text{2D}}}\right)\exp{\left(\frac{-d_{\text{2D}}}{p_1}\right)}+ \frac{d_1}{d_{\text{2D}}} ,\quad d_{\text{2D}} > d_1.
        \end{cases}
    \end{equation}
    Where $d_\text{2D}$ represents the horizontal distance between the aerial vehicle and the ground base station, and $h$ denotes the height. Meanwhile, $d_1$ and $p_1$ are defined as:
    \begin{equation}\label{eq:d1p1}\begin{split}
        d_1 =& \max(294.05\log_{10}(h) - 432.94, 18);\\
        p_1 =& 233.98\log_{10}(h) - 0.95.\\
    \end{split}
    \end{equation}
    Summarizing \gls{los} and \gls{nlos} conditions, the average path loss component can be described with, 
    \begin{equation}\begin{split}
        \eta =& (4.32 - 0.76\log_{10}(h))(1-P_\text{los}) \\
        &+  (2.225 - 0.05\log_{10}(h))P_\text{los}
    \end{split}
   \end{equation}
    When $d_{\text{2D}} \leq d_1$, the channel is dominated by \gls{los}, which we define as area 1 (A1). Conversely, when $d_{\text{2D}} > d_1$, the channel can be either \gls{los} or \gls{nlos} depending on the distance and height; consequently, we designate this region as area 2 (A2). We then formulated the average $\eta$ in Eq.~(\ref{eq:eta1}), the first derivative of $\eta$ with respect to $d_{\text{2D}}$ in Eq.~(\ref{eq:etaderia}), and the second derivative in Eq.~(\ref{eq:etasecderia}), respectively.
   \begin{align}\label{eq:eta1}
   \eta =
   \begin{cases}\begin{split}
     2.225& - 0.05\log_{10}h,\quad \text{if in A1};\\
    (4.32 &- 0.76\log_{10}(h))(1-P_\text{los}) \\
        &+  (2.225 - 0.05\log_{10}(h))P_\text{los},\quad \text{if in A2}.
     \end{split}
   \end{cases}
   \end{align}
   \begin{align}\label{eq:etaderia}
   \eta' =
   \begin{cases}\begin{split}
     &0,\quad \text{if in A1};\\
     &\bigg(\underbrace{0.71\log_{10}(h) - 2.07}_{<0}\bigg) \Bigg(\Bigg[\underbrace{-\frac{d_1}{d^2_{\text{2D}}}- 
      \frac{1 - \frac{d_1}{d_{\text{2D}}}}{p_1} }_{<0}\Bigg]\\
     & \quad\quad\underbrace{\exp\left(-\frac{d_{\text{2D}}}{p_1}\right)}_{>0} - \underbrace{\frac{d_1}{d^2_{\text{2D}}}}_{>0}\Bigg)
   \quad \text{if in A2}.
     \end{split}
   \end{cases}
   \end{align}
   
    \begin{align}\label{eq:etasecderia}
   \eta'' =
   \begin{cases}\begin{split}
    &0,\quad \text{if in A1};\\  
    &\bigg(\underbrace{0.71\log_{10}(h) - 2.07}_{<0}\bigg)\bigg(\bigg[\underbrace{\frac{2d_1}{d_{\text{2D}}^3} + \frac{d_{\text{2D}}-d_1}{p^2_1 d_{\text{2D}}}\bigg]}_{>0} \\
    &\quad\quad\underbrace{\exp\left(-\frac{d_{\text{2D}}}{p_1}\right)}_{>0} + \underbrace{\frac{2d_1}{d_{\text{2D}}^3}}_{>0}\bigg) \quad \text{if in A2}.   
     \end{split}
   \end{cases}
   \end{align}
   Considering the practical constraints, the \gls{uav} must fly above a minimum height to avoid collisions in urban areas while remaining below a maximum height to comply with legal regulations \cite{bmvd2021drones}, we have:
   \begin{equation}\label{eq:c1}
   \mathrm{C1}: h \in [20,120].
   \end{equation}
   Under constraint $\mathrm{C1}$, we observe that $0.71\log_{10}(h) - 2.07 < 0$, therefore both $\eta' \geq 0$ and $\eta'' \leq 0$ hold. This indicates that $\eta$ is monotonically increasing with diminishing rate. The averaged values of $\eta$ are presented in Fig.~\ref{fig:a2gchannel}. It should be noted that the \gls{a2g} model, according to the definition in \cite{3gpp-tr36.777}, is only applicable when $h \geq \SI{22.5}{\meter}$. Below $\SI{22.5}{\meter}$, a different model from \cite{3gpp.38.901} should be applied. However, to simplify our analysis, we extended the \gls{a2g} model down to $\SI{20}{\meter}$.
	\section{Problem formulation}\label{sec:sys_model}
    \subsection{Objective equivalence}
    \gls{crb} of \gls{3d} localization can be described as a formula of the modeled distance error $\sigma_{M}$ and the number of reference nodes $N$,
    
    \begin{equation}\label{eq:loccrb}
        \sigma^2_{\theta} \ge \frac{6\sigma^2_{M}}{N}
    \end{equation}
    Eq.~(\ref{eq:loccrb}) is derived under the assumption that all reference nodes are uniformly distributed around the \gls{uav}, with numerical validation provided in \cite{ZYGMsecureUAV2018}. While $\sigma_{M}$ comprises both position error variance $\sigma_p$ and distance error variance $\sigma_d$, we can simplify our analysis by considering static references as reliable. Despite neglecting $\sigma_p$, $\sigma_{M}$ remains mathematically intractable. Therefore, we simplify it to a tractable approximation:
    \begin{equation}\label{eq:loccrb2}
    {\sigma}^2_{\theta} \propto \frac{1}{N}\big(\frac{\sum_{n=1}^N\sigma_{n,d}}{N}\big)^2,
   \end{equation}
    where ${\sigma}_{n,d}$ denotes the distance error variance for individual communication links with nodes. According to \cite{Venus2020ToA, Hechen2024ToA}, under dense multi-path scenario the \gls{crb} for \gls{tdoa} measurements is $\sigma_d^2\geq J_{\text{T}}^{-1}$
    , where $J_{\text{T}}$ represents the Fisher information for \gls{tdoa} estimation, with
    \begin{equation}\label{eq:toacrb}
    J_{\text{T}} = {2 c^{-1} 4 \pi^2 \text{SINR}\gamma \beta^2 \sin^2(\phi)}.
    \end{equation}
    where $c$ is the speed of light and $\text{SINR}$ denotes the \gls{sinr}, with interference primarily due to multipath effects. Since the \gls{uav}'s sub-channels have closely spaced carrier frequencies in a dense urban environment, their multipath profiles are highly correlated and vary predominantly with altitude. Thus, we model $\text{SINR} = \text{SNR} \cdot R$, where $R$ represents an interference-to-noise ratio factor treated as constant across links to simplify analysis.
    Meanwhile, the parameters $\gamma$ and $\sin^2(\phi)$ represent the whitening gain and information loss due to path loss nuisance parameters, respectively. As demonstrated in \cite{WGKlaus2016}, while both $\gamma$ and $\sin^2(\phi)$ vary with $\beta$, the variation in $\sin^2(\phi)$ is negligible compared to that of $\gamma$. Consequently, $J_{\text{T}}$ is predominantly scaled by $\text{SNR}$, $\beta$, and $\gamma$. The whitening gain $\gamma$ can be characterized with,
    \begin{equation}\nonumber
    \gamma = \frac{\text{SNR}_\text{w}}{\text{SNR}} = \frac{P_{\text{wpre}}}{P_{\text{wpost}}}
    \end{equation}
    \begin{equation}\nonumber
    P_{\text{wpre}} = \int_{-\frac{\beta}{2}}^{\frac{\beta}{2}} s_n(f) df; \quad P_{\text{wpost}} = \text{S}_n^w\beta.
    \end{equation}
    Where $P_{\text{wpre}}$ and $P_{\text{wpost}}$ denote the noise power before and after the whitening process, After whitening, the PSD of noise become flat, denoted as $\text{S}_n^w$. Consider $s_n(f)$ to be modeled with power-law noise model $s_n(f) = C f^{-\alpha}$. Then we can have, 
     \begin{equation}\nonumber\begin{split}
       P_{\text{wpre}} &= \int_{-\frac{\beta}{2}}^{\frac{\beta}{2}} C f^{-\alpha} df =2\alpha C\ln(\frac{\beta}{2}).
     \end{split} 
     \end{equation}
    Considering $\alpha = 1$ is common in real-world systems, therefore,
    \begin{equation}\nonumber
     \gamma = \frac{P_{\text{wpre}}}{P_{\text{wpost}}} = \frac{2C\ln(\frac{\beta}{2})}{\text{S}_n^w\beta},
    \end{equation}
    we can further simplify this to $\gamma \propto \ln(\beta)\beta^{-1}$, since $\beta$ is the bandwidth, typically $\ln(\beta)\gg\ln(2)$. The channel between \glspl{uav} and ground infrastructures can be modeled as a Rician channel. The $K$ factor also influences many aforementioned factors in Eq.~(\ref{eq:toacrb}). In urban environments, the $K$ factor typically exhibits limited variation subjected to the altitude; therefore, subsuming $\gamma$ into Eq.~(\ref{eq:toacrb}), we reasonably assume that $\sigma^2_{n,d}$ is primarily governed by two factors, with the proportional relationship: $\sigma^2_{n,d} \propto(\text{SNR}_n\beta_n\ln\beta_n)^{-1}$. Considering that in practice, transmitting power is usually proportional to bandwidth, and so is the noise floor,
    \begin{equation}\label{eq:transp}
    P_n = \psi \beta_n, \quad N_n = N_0 \beta_n,
    \end{equation}
    where $\psi$ is the transmitting power spectral density and $N_0$ is the noise power spectral density, respectively. Since sub-band bandwidths are typically much smaller than the central carrier frequency, differences in carrier frequency can be neglected. 
    Under these assumptions, the proportion $\text{SNR}_n \propto d_{\text{3D},n}^{-\eta_n}$ holds. With $\sigma^2_{n} \propto d_{\text{3D},n}^{\eta_n}(\beta_n\ln\beta_n)^{-1}$, substituting this into Eq.~(\ref{eq:loccrb2}):
    \begin{equation}\label{eq:locobj}
    {\sigma}^2_{\theta} \propto {N}^{-3}\left(\sum_{n=1}^N d_{\text{3D},n}^{\frac{1}{2}\eta_n}(\beta_n\ln\beta_n)^{-\frac{1}{2}}\right)^2.
   \end{equation}
  
   \subsection{Reference deployment}
\begin{figure}[!htbp]
    \centering
    \includegraphics[width=0.85\linewidth]{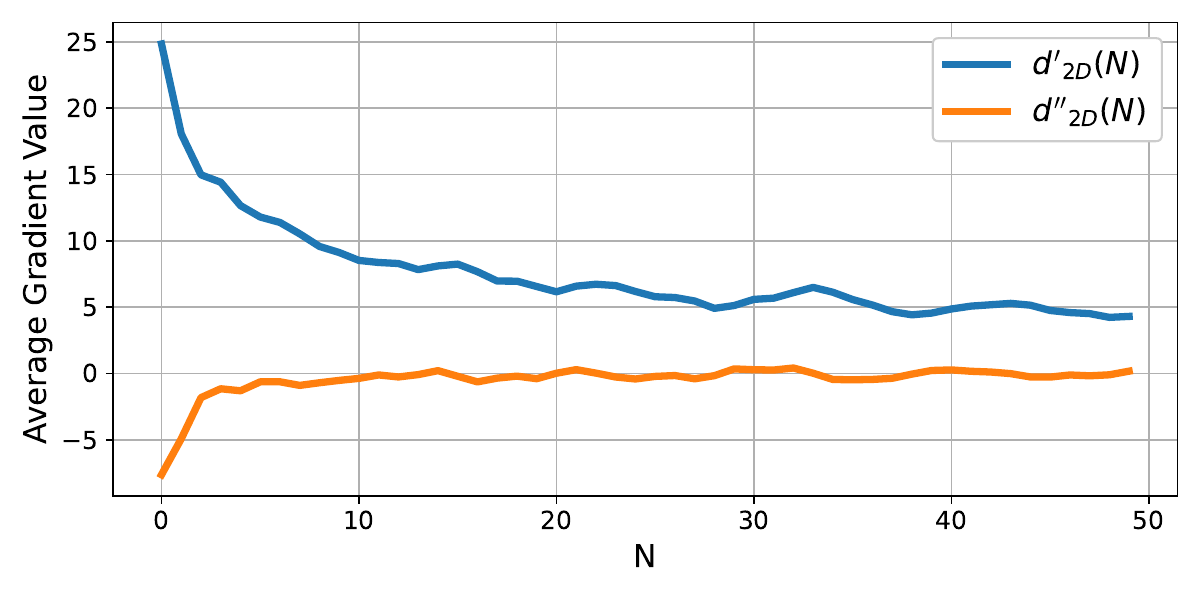}
    \caption{$d'_{\text{2D}}(N)$ and $d''_{\text{2D}}(N)$ when $R=\SI{120}{\meter}$}\label{fig:d2dd}
    \end{figure}

   When an \gls{uav} is flying above a dense urban area, it is reasonable to assume that reference nodes from different directions are available. In such cases, the \gls{uav} can select the nearest reference nodes to optimize its self-localization performance. Therefore, the $d_\text{2D}$ to $N_\text{th}$ nearest node can be described as a monotonically increasing function $d_{\text{2D}}(N)$.   
   \begin{figure}[!ht]
    \centering
    \includegraphics[width=0.69\linewidth]{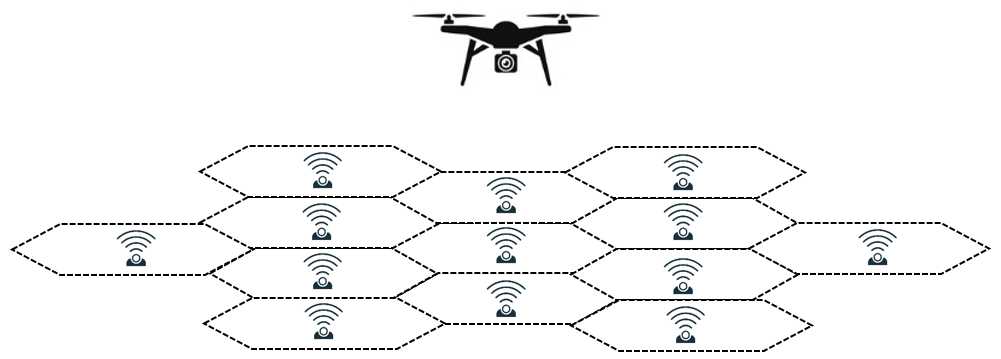}
    \caption{system model}\label{fig:sysmodel}
    \end{figure}
  For a hexagonal grid topology for reference deployment, depicted in Fig.~\ref{fig:sysmodel}. $d_{\text{2D}}(N)$ is:
    \begin{equation}\label{eq:Dnaps}
    d_{\text{2D}} (N)
    \begin{cases}
    =\Delta, 
    \quad \text{if } N = 1, \\
    \in [R - \Delta, R + \Delta],
    \quad \text{if } 1 < N \leq 7, \\
    \vdots \\
    \in [kR - \Delta, kR + \Delta], \\
    \quad \text{if } 3k^2 - 3k + 1 < N \leq 3k^2 + 3k + 1,
   \end{cases}
   \end{equation}
    where $\Delta$ is the distance to the $1^\text{st}$ reference node, $R$ is the coverage of reference nodes, and $\Delta\in [0,R/2]$. $k$ is the layer of the hexagonal grids. Apparently, $d_{\text{2D}}(N)$ is a piecewise function; however, we can still summarize its derivative segmentwise as,
    \begin{equation}\label{eq:Dnaps2}
    d'_{\text{2D}}(N)  
    \begin{cases}
    =\Delta, & N = 1 ,\\
    \approx\frac{\Delta}{3}, &1 < N \leq 7,\\
    \vdots\\
    \approx\frac{\Delta}{3k}, &3k^2 - 3k + 1 < N \leq 3k^2 + 3k + 1.
    \end{cases}
    \end{equation}
    We can conclude that, $d'_{\text{2D}}(N)>0$ and $d''_{\text{2D}}(N) \approx 0$ in most cases, with averaged values of $5000$ numerical simulation depicted in Fig.~\ref{fig:d2dd}. 

   \subsection{Localization optimization}
   In multi-\gls{uav} systems sharing ground infrastructure, subband bandwidth can be fixed and pre-determined for efficient multiplexing. We analyze localization performance when $\beta_n$ remains constant, where each \gls{uav} optimizes its own localization accuracy based on available information. Consequently, Eq.~(\ref{eq:locobj}) can be further simplified as following,
   \begin{equation}\label{eq:locobj1}
    {\sigma}^2_{\theta} \propto {N}^{-3}\left(\sum_{n=1}^N d_{\text{3D},n}^{\frac{1}{2}\eta_n} \right)^2,
   \end{equation}
    \begin{equation}\label{eq:locobj2}
   \sum_{n=1}^N d_{\text{3D},n}^{\frac{1}{2}\eta_n} = \sum_{n=1}^N (d_{\text{2D},n}^2 + h^2)^{\frac{1}{4}\eta(d_{\text{2D},n}, h)},
   \end{equation}
   \begin{equation}\label{eq:capphid2dh}
     \Phi(N) =  \sum_{n=1}^N (d_{\text{2D},n}^2 + h^2)^{\frac{1}{4}\eta(d_{\text{2D},n}, h)} ,
   \end{equation}
   \begin{equation}\label{eq:phid2dh}
     \phi(N) = (d_{\text{2D},n}^2 + h^2)^{\frac{1}{4}\eta(d_{\text{2D},n}, h)},
   \end{equation}
   and referring Eq.~(\ref{eq:locobj1}), we formulate the optimization problem: 
   \begin{equation} \label{eq:opt_problem}
  \begin{aligned}
   \underset{{N}}{\text{min}} \quad &  F_\theta(N) = \Phi^2(N)N^{-3} \\
   \text{s. t.} :\quad & N \in \mathbb{Z}_{>1}, \\
                       & h \in \mathbb{R}^{+},\\
                       & d_{\text{2D},n} \in \mathbb{R}^{+}, \quad d_{\text{2D},n} \geq \Delta,\quad n \in \mathcal{N}.
   \end{aligned}
   \end{equation}
   $\Phi(N)$ is discrete sum function and can be approximated by continuous function by applying the Euler–Maclaurin summation formula: 
   \begin{equation}
   \Phi(N) \approx \int_1^N \phi(x)\,dx + \frac{\phi(N)}{2} + \frac{\phi'(N)}{12} - \frac{\phi'''(N)}{720} + \cdots
   \end{equation}
   Correspondingly, $\Phi'(N)$ takes the form,
   \begin{equation}
   \Phi'(N) \approx \phi(N) + \frac{\phi'(N)}{2} + \frac{\phi'''(N)}{12} - \frac{\phi''''(N)}{720} + \cdots
   \end{equation}
   The first-order derivative and the second-order derivative of $\phi(N)$ are given in Eqs.~(\ref{eq:1stderiaN})--(\ref{eq:2edderiaN}). Recalling Eqs.~(\ref{eq:eta1})--(\ref{eq:etasecderia}), our numerical results indicate mean values for $\eta$, $\eta'$, and $\eta''$ of $2.3$, $4.8 \times 10^{-3}$, and $2.1 \times 10^{-5}$, respectively. We can conclude that $\phi(N) \gg \frac{\phi'(N)}{2} \gg \frac{\phi''(N)}{12}$. To simplify our analysis, we take:
   \begin{equation}
   \Phi'(N) \approx \phi(N) + \frac{\phi'(N)}{2}
   \end{equation}
    \begin{figure*}[!t]
    \begin{align}\label{eq:1stderiaN}\begin{split}
        \phi'(N) = \frac{\partial\phi(N)}{\partial d_{\text{2D}}}\frac{\partial d_{\text{2D}}}{\partial N}  =\underbrace{\left( d_{\text{2D}}^2 + h^2 \right)^{\frac{1}{4} \eta} \left[ \frac{1}{4} \cdot  \eta' \cdot \ln(d_{\text{2D}}^2 + h^2) + \frac{2d_{\text{2D}}\eta}{d_{\text{2D}}^2 + h^2} \right](d'_{\text{2D}})}_{>0}
    \end{split}
    \end{align}
    \begin{equation}\label{eq:2edderiaN}
    \begin{split}
     \phi''(N) &=\frac{\partial^2\phi(N)}{\partial^2 d_{\text{2D}}}\bigg(\frac{\partial d_{\text{2D}}}{\partial N}\bigg)^2+\frac{\partial\phi(N)}{\partial d_{\text{2D}}}\frac{\partial^2 d_{\text{2D}}}{\partial^2 N}\\
     &= \left( d_{\text{2D}}^2 + h^2 \right)^{\frac{1}{4} \eta} \Bigg[
     \underbrace{\frac{\ln(d_{\text{2D}}^2 + h^2)}{16} \bigg((\eta')^2 \ln(d_{\text{2D}}^2 + h^2) - \frac{1}{4} \eta''\bigg)}_{>0} + \underbrace{\frac{d_{\text{2D}}\eta\eta'\ln(d_{\text{2D}}^2 + h^2)+2 \eta  + 2.5d_{\text{2D}} \eta'}{d_{\text{2D}}^2 + h^2}}_{>0}\Bigg](d'_{\text{2D}})^2 \\  &+ \underbrace{\left( d_{\text{2D}}^2 + h^2 \right)^{\frac{1}{4} \eta} \left[     \frac{1}{4} \eta' \cdot \ln(d_{\text{2D}}^2 + h^2) +     \frac{2d_{\text{2D}}\eta}{d_{\text{2D}}^2 + h^2} \right](d''_{\text{2D}})}_{\approx0}   
    \end{split}
    \end{equation}
   \begin{align}\label{eq:deriaNobj}
   \begin{split}
   F'_\theta(N) &\approx -3N^{-4}\Phi^2(N) + 2N^{-3}\Phi(N)\Phi'(N) = \underbrace{-3N^{-4}\Phi^2(N)}_{<0} + \underbrace{2N^{-3}\Phi(N)\left( \phi(N)+\frac{\phi'(N)}{2}\right)}_{>0}
  \end{split}
  \end{align}
   \begin{align}\label{eq:secderiaNobj}
   \begin{split}
   F''_\theta(N) &= 12N^{-5} \Phi^2(N) - 12N^{-4} \Phi(N)\Phi'(N)+2N^{-3}\bigg((\Phi'(N))^2+\Phi(N)\Phi''(N)\bigg)\\
   &= \underbrace{12N^{-5} \Phi^2(N)}_{>0} - \underbrace{12N^{-4} \Phi(N)\left( \phi(N) + \frac{\phi'(N)}{2}  \right)}_{>0}  + \underbrace{2N^{-3} \left[ \left( \phi(N) + \frac{\phi'(N)}{2}  \right)^2 + \Phi(N)\left( \phi'(N) + \frac{\phi''(N)}{2}  \right) \right]}_{>0}
  \end{split}
  \end{align}
  \rule{\textwidth}{0.4pt} 
   \vspace*{\fill} 
  \end{figure*}
   Meanwhile, the first-order derivative and the second-order derivative of $F_\theta(N)$ are given in Eqs.~(\ref{eq:deriaNobj})--(\ref{eq:secderiaNobj}). With the given information, we cannot determine the monotonicity of $F_\theta(N)$. Furthermore, the sign of $F''_\theta(N)$ is difficult to determine since both positive and negative terms are involved. However, we can write $F'_\theta(N)$ in the form: 
   \begin{align}\label{eq:deriaN2}
   F'_\theta(N) &= \underbrace{2N^{-3}\Phi(N)}_{T_1(N)}\underbrace{\bigg( \phi(N) - \frac{3}{2N}\Phi(N) + \frac{\phi'(N)}{2}\bigg)}_{T_2(N)},
  \end{align}
  \begin{align}\label{eq:SecderiaN}
   \frac{\partial T_2(N)}{\partial N} = (\frac{2N-3}{2N})\phi'(N)+\frac{3}{2N^2}\phi(N)+\frac{1}{2}\phi''(N).
  \end{align}
   Within Eq.~(\ref{eq:deriaN2}), clearly $T_1(N) > 0$, and the sign of $T_2(N)$ can't be determined. Consequently, $\frac{\partial T_2(N)}{\partial N}$ is given in Eq.~(\ref{eq:SecderiaN}). In this equation, while $N\geq2$, $\frac{\partial T_2(N)}{\partial N} \geq 0$ always holds; in the case $N = 1$, we still have $\frac{3\phi(N)}{2} > \frac{\phi'(N)}{2}$. Therefore, $\frac{\partial T_2(N)}{\partial N} > 0$ always holds, indicating that $T_2$ is monotonically increasing. When $N = 1$, $T_2(N) = -\frac{\phi(N)}{2} + \frac{\phi'(N)}{2}$. With $\phi(N) \gg \phi'(N)$ and $\lim_{N\to\infty}T_2(N) > 0$, we can conclude that $T_2(N)$ is not only monotonically increasing, but also increasing from a negative value to a positive value. In summary of all analysis, despite the lack of convexity proof for $F_\theta(N)$, at the point where $T_2(N) = 0$, its left side satisfies $F'(\theta) < 0$ and its right side satisfies $F'(\theta) > 0$, therefore an optimal solution exists.
   \section{proposed method}\label{sec:method}
   Facilitating the average \gls{a2g} channel statistics, previous analysis demonstrated an optimal node selection number $N_\text{opt}$ exists. However, in practice, $\eta$ of different links cannot be obtained without extensive channel estimation. Meanwhile, accessing $d_{\text{2D}}(N)$ is challenging since ranging methods are error-prone, and using \gls{tdoa} for all communication links is computationally expensive and introduces latency. Since precisely solving $N_\text{opt}$ introduces significant complexity, we propose a lightweight approach using only \gls{rssi}, offering advantages: \begin{enumerate*}[label=\emph{\roman*)}]
   \item extremely lightweight, requiring no additional processing;
   \item nodes with poor channels are automatically excluded due to significantly large estimated distances;
   \item sensitive to \gls{los} probability mismatch caused by environmental factors in \gls{a2g} models, enabling possible compensation
   \end{enumerate*}. The detailed algorithm is introduced in Alg.~\ref{alg:rbof}.
   
    After assessing the optimal strategy for localization, \gls{tdoa} of selected nodes will be measured, we then employ the gradient algorithm presented in \cite{GD2025fang} for localization. This algorithm is superior to other localization algorithms in scenarios where the reference positions exhibit limited dynamic range in altitude. 
    \begin{algorithm}[!htbp]
    \caption{RSSI-based optimum finder}
    \label{alg:rbof}
    \scriptsize
    \DontPrintSemicolon
    Input: estimated distances based on \gls{rssi} $d^R_{n}$; altitude $h$; averaged path loss component consolidated as a table $\overline\eta(d_\text{2D}, h)$; maximum reference numbers $N_m$ \\
    \SetKwProg{Fn}{Function}{ :}{end}
    \Fn{}{ 
        sort $d^R_{n}$; \\
        \For {$n = 1:N_m$ }{
        $d_{\text{2D},n} = \sqrt{(d^R_{n})^2 - h^2}$\\
        find $\overline\eta$ regarding $d_{\text{2D},n}$ and $h$\\
        compute $\phi_n$ referring Eq.~(\ref{eq:phid2dh})
        }
        compute $\Phi_n$ referring Eq.~(\ref{eq:capphid2dh})\\
        compute $T_{2,n}$ referring Eq.~(\ref{eq:deriaN2})\\
        $T_2 \gets \{T_{2,n}; n \in [1, N_m]\}$\\
       $ N_\text{opt} \gets \left| \left\{n \in [1, N_m]  \;:\; T_2[n] < 0 \right\} \right|$ \tcp{Count the number of negative values}
       \If{$\overline T_2 > 0 $}{$T^\circ_2= T_2- \overline T_2$}
       \Else{$T^\circ_2= T_2+ |\overline T_2|$\tcp{compensate for $\eta$ miss-match}}
       $ N^\circ_\text{opt} \gets \left| \left\{n \in [1, N_m]  \;:\; T^\circ_2[n] < 0 \right\} \right|$\\
       $N_\text{opt} = \operatorname{round}(\frac{N_\text{opt}}{2} + \frac{N^\circ_\text{opt}}{2})$ \\
       $N_\text{opt} = \min(20,\max(N_\text{opt},3))$ \tcp{set upper and lower boundaries}
       }
    \end{algorithm}
 
   \section{Simulation results}\label{sec:simu}
\begin{figure}[!htbp]
    \centering
    \begin{subfigure}[b]{0.40\textwidth}
        \centering
        \includegraphics[width=\linewidth]{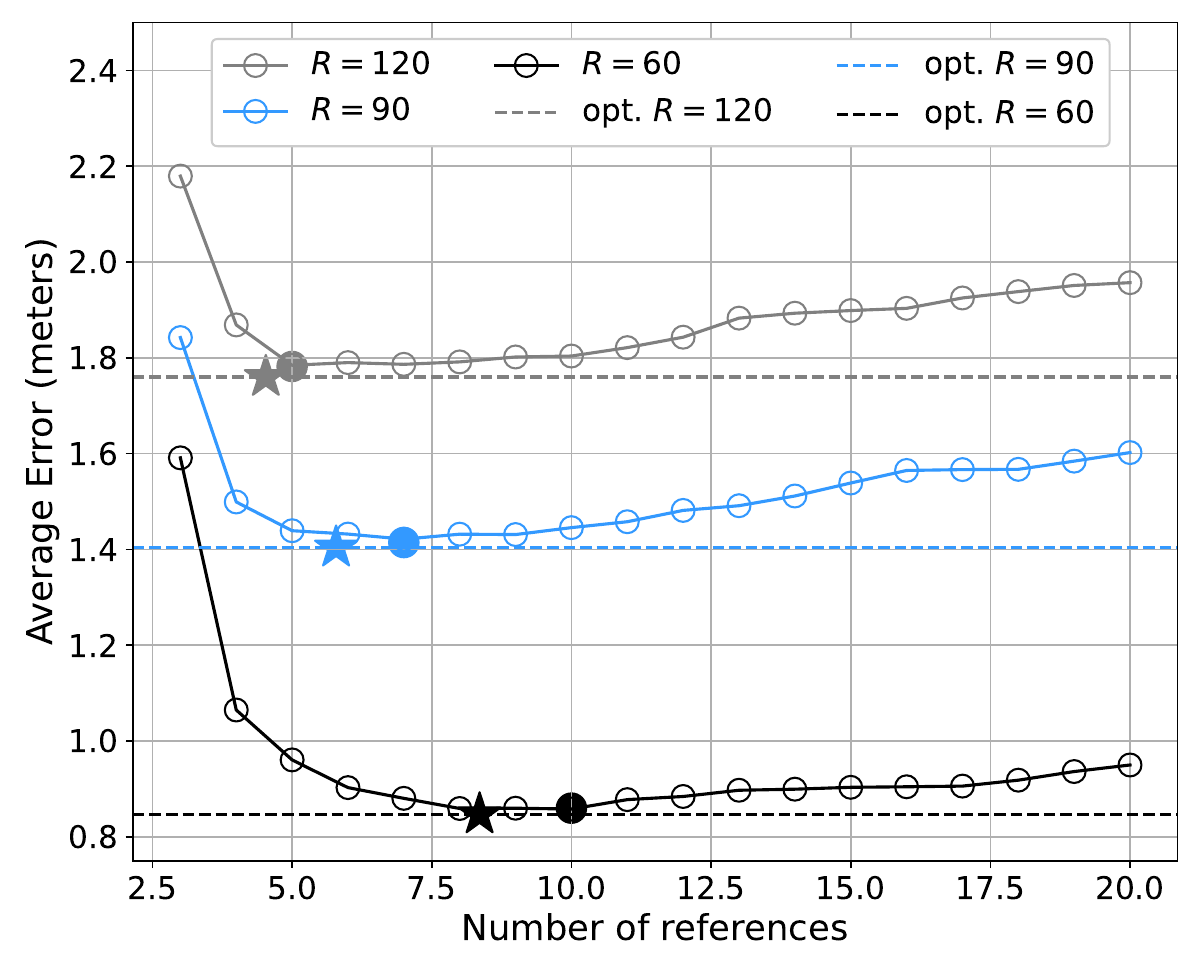}
    \end{subfigure}
    \begin{subfigure}[b]{0.40\textwidth}
        \centering
        \includegraphics[width=\linewidth]{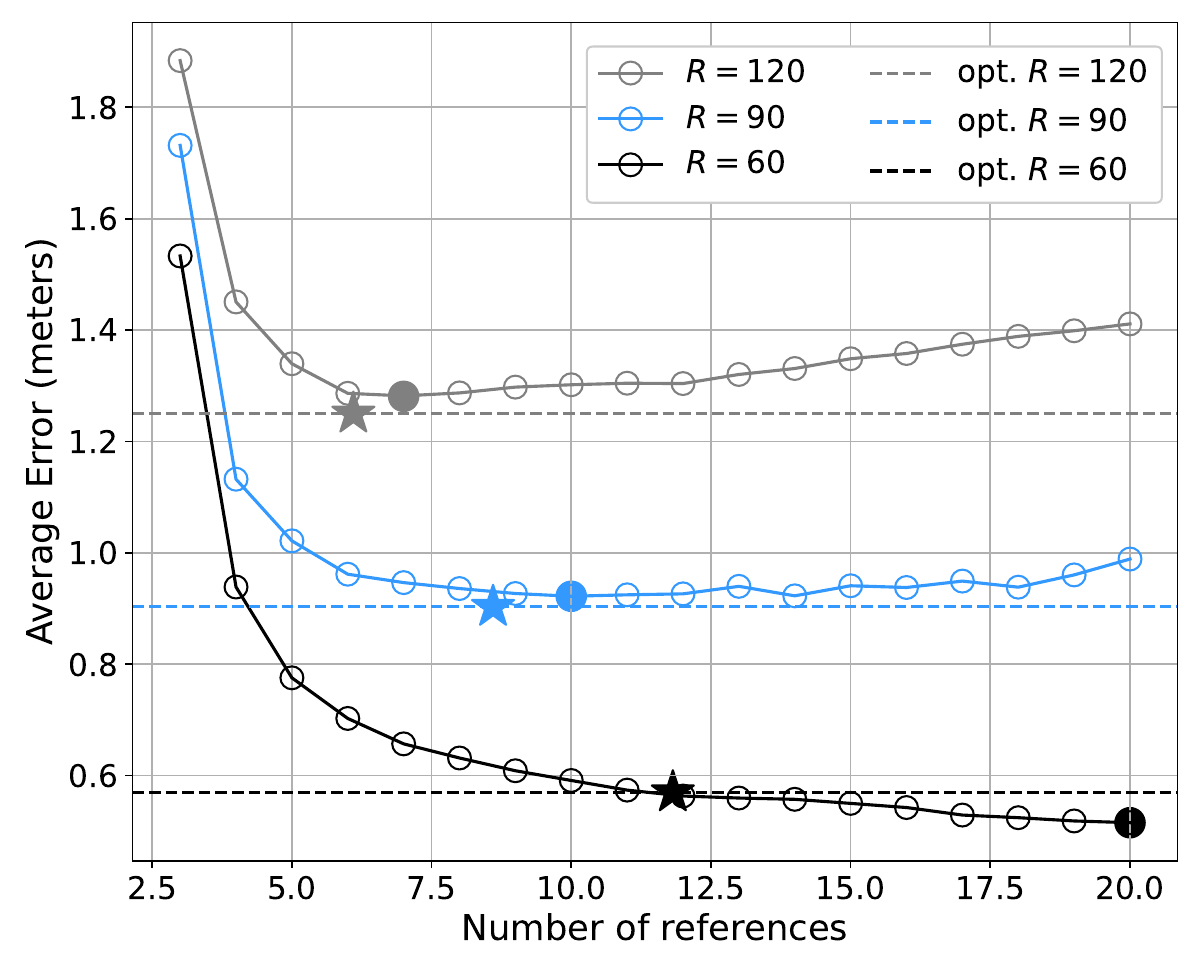}
    \end{subfigure}\caption{Localization performance with $h_u = \SI{20}{\meter}$ (top plot) and $h_u = \SI{30}{\meter}$ (bottom plot).}\label{fig:optsimresu}
    \end{figure}

   In this section, we validate the performance of Alg.~\ref{alg:rbof} under different deployment configurations of reference nodes and varying altitudes of the \gls{uav}. We assume the \gls{uav} is positioned in the relative center of the area with reference nodes evenly distributed from all directions, eliminating unilateral reference problems. All sensor nodes are synchronized and periodically broadcast equal-powered reference signals containing their positions and \gls{tdoa} information. The detailed simulation setup is listed in Tab.\ref{tab:setup1}.
      \begin{table}[!htbp]
		\centering
        \scriptsize
		\caption{Simulation setup 1}
		\label{tab:setup1}
		\begin{tabular}{>{}m{0.2cm} | m{1.6cm} l m{3.7cm}}
			\toprule[2px]
			&\textbf{Parameter}&\textbf{Value}&\textbf{Remark}\\
			\midrule[1px]        
			&$f_c$&$3.5$ Ghz& Carrier frequency\\

			&$K$&$(0.1,3.0)$& Rician factors\\
			&$N_p$&$4$& Number of multipath\\
            & $\tau_\text{max}$ & 2e-7 s &  Maximum delay spread\\ 
            & $P_\text{t}$ & 15 dBm &  Transmitting power\\ 
            & $N_\text{o}$ & -91 dBm &  Noise floor\\ 
            & $\beta_n$ & 10 Mhz &  Bandwidth\\ 
            & $\sigma_t$ & $1 \mu s $&  Average synchronization error\\
            \midrule[1px]
            \multirow{-9.9}{*}{\rotatebox{90}{\textbf{TDOA}}}
            & $h_u $&$[20,30]$& \gls{uav} altitude\\ 
            &$\sigma_h$&$1$& Altitude measurement error power\\
			&$\sigma_\text{GPS}$&$5$& Initial GPS error power\\
            &$R$&$[60,90,120]$& Node coverage\\
            &$h_n$&$\sim\mathcal{U}(0,5)$& Node Altitudes\\
			\multirow{-5.2}{*}{\rotatebox{90}{\textbf{Deployment}}}
            & $\Delta_\text{LOS}$ & $[-0.4,0.1]$&  \gls{los} probability modification\\
            \bottomrule[2px]
		\end{tabular}
	\end{table}
   Firstly, we evaluate the performance of Alg.~\ref{alg:rbof}. We limit the reference number within a range of $3$ to $20$. To assess \gls{3d} coordinates, at least $3$ references are required, while involving too many reference nodes is impractical in practice. Therefore, we limit the reference numbers to $20$. The consolidated results of $1000$ simulations are presented in Fig.~\ref{fig:optsimresu}.
  
   The simulation results confirm our analytical findings: localization considering node distribution has an optimal solution, and its solution $N_\text{opt}$ varies with respect to node coverage and altitude. Despite $h_u = \SI{30}{\meter}$ resulting in generally longer distances to the sensor nodes, the localization performance is still superior due to improved channels. The localization performance when applying Alg.~\ref{alg:rbof} demonstrates a slight enhancement in localization accuracy and requires fewer reference nodes by providing a dynamic solution to different circumstances. Except for the case where $h_u = \SI{30}{\meter}$ and $R = \SI{60}{\meter}$, the average optimal solution is apparently beyond $20$. Under this configuration, Alg.~\ref{alg:rbof} fails to provide comparable performance to $N=20$. The true $N_\text{opt}$ is frequently larger than $20$. When applying Alg.~\ref{alg:rbof}, $N_\text{opt}$ is bounded by $20$, which consequently restricts performance. However, when significantly enlarging $N$, the performance enhancement is rather limited. Notably, the localization performance in this case is already satisfactory.
    
\begin{figure}[!htbp]
    \centering
    \begin{subfigure}[b]{0.40\textwidth}
        \centering
        \includegraphics[width=\linewidth]{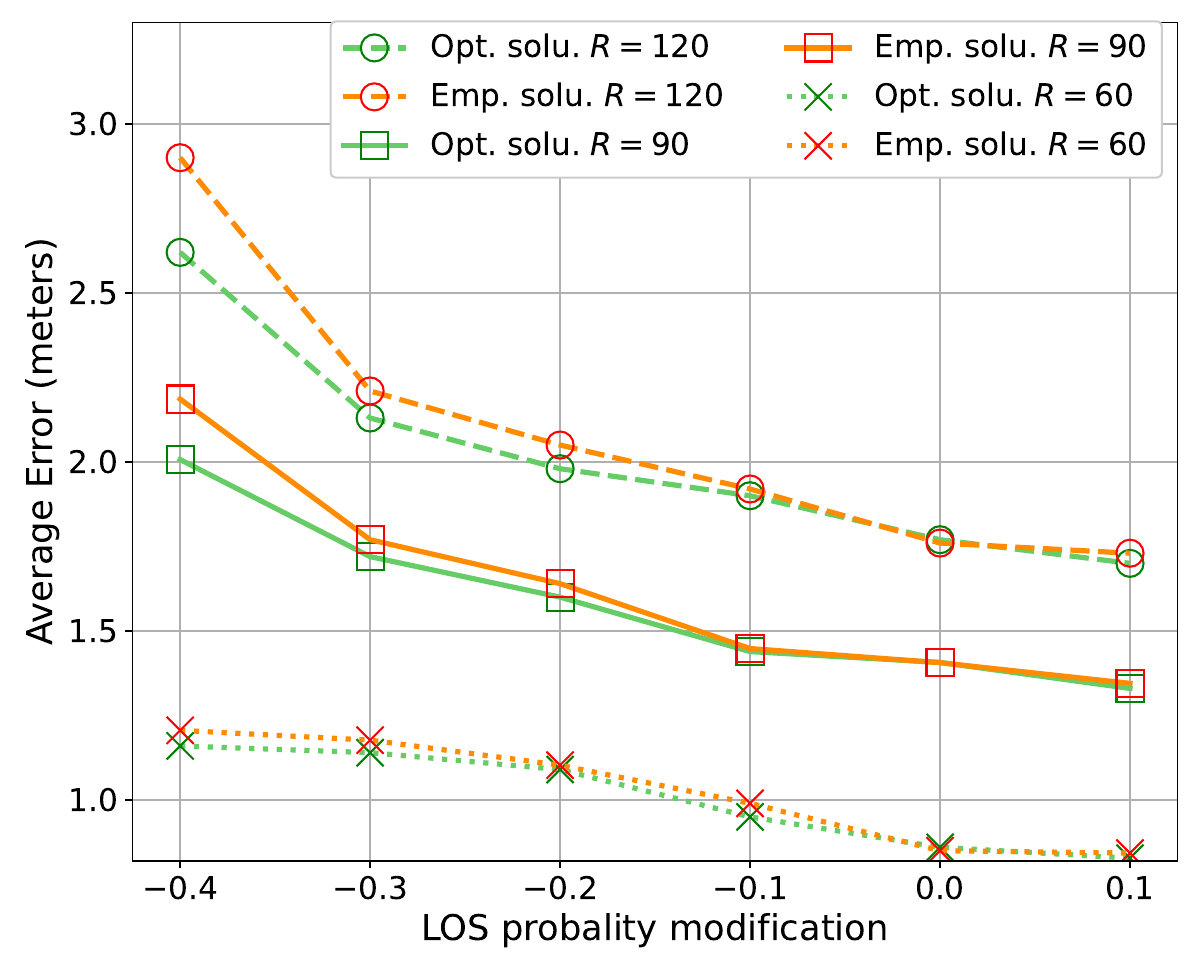}
    \end{subfigure}
    \begin{subfigure}[b]{0.40\textwidth}
        \centering
        \includegraphics[width=\linewidth]{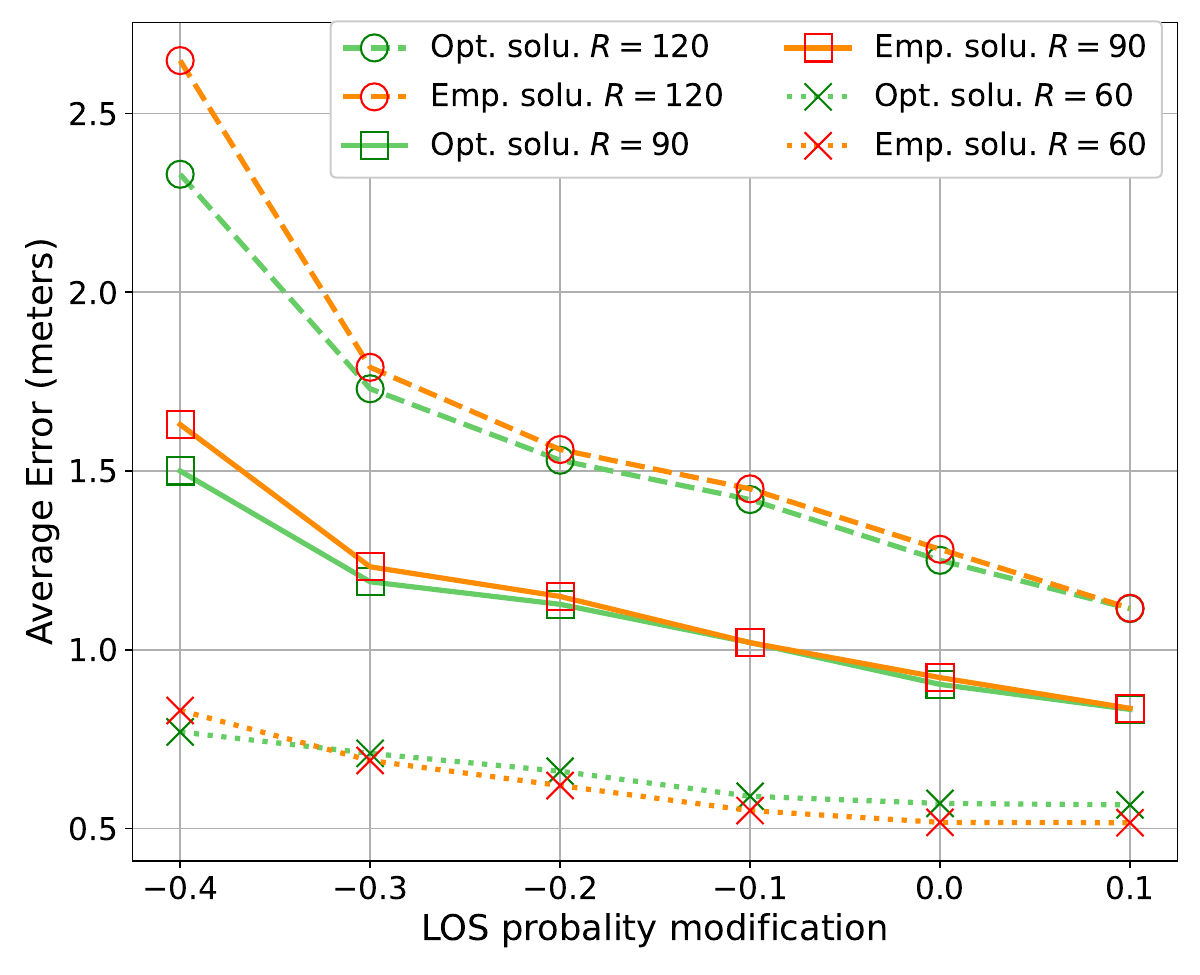}
    \end{subfigure}\caption{Localization performance with \gls{los} modification  $h_u = \SI{20}{\meter}$ (top plot) and $h_u = \SI{30}{\meter}$ (bottom plot).}\label{fig:lossimresu}
    \end{figure}

    It is reasonable to assume that when the \gls{uav} precisely knows its altitude and the deployment of ground reference nodes, an empirical optimal solution can be applied for simplicity. However, even neglecting altitude estimation errors and deployment knowledge ambiguity, \gls{los} probability can be heavily impacted by environmental factors, making such empirical solutions infeasible. We therefore modify the $P_\text{los}$ on top of Eq.(\ref{eq:problos}) and compare the empirical solution with the optimization solution based on Alg.\ref{alg:rbof}. The simulation results in Fig.~\ref{fig:lossimresu} demonstrate that the optimization solution consistently outperforms the empirical solution, particularly under scenarios with poor \gls{los} conditions, showing resilience to \gls{los} variations.
   
   \section{Conclusion}\label{sec:conclu}
   Our study presents a resource-efficient framework for \gls{uav} self-localization by optimizing reference node selection under practical constraints. Simulations validate that the proposed method enhances accuracy while adapting to varying channel conditions. Future work will explore real-world considerations such as unilateral reference problems and dynamic network synchronization for scalable \gls{uav} localization in \gls{6g} networks.

\bibliographystyle{IEEEtran}
\bibliography{references}

\begin{thebibliography}{10}
\providecommand{\url}[1]{#1}
\csname url@samestyle\endcsname
\providecommand{\newblock}{\relax}
\providecommand{\bibinfo}[2]{#2}
\providecommand{\BIBentrySTDinterwordspacing}{\spaceskip=0pt\relax}
\providecommand{\BIBentryALTinterwordstretchfactor}{4}
\providecommand{\BIBentryALTinterwordspacing}{\spaceskip=\fontdimen2\font plus
\BIBentryALTinterwordstretchfactor\fontdimen3\font minus
  \fontdimen4\font\relax}
\providecommand{\BIBforeignlanguage}[2]{{%
\expandafter\ifx\csname l@#1\endcsname\relax
\typeout{** WARNING: IEEEtran.bst: No hyphenation pattern has been}%
\typeout{** loaded for the language `#1'. Using the pattern for}%
\typeout{** the default language instead.}%
\else
\language=\csname l@#1\endcsname
\fi
#2}}
\providecommand{\BIBdecl}{\relax}
\BIBdecl

\bibitem{LAE2025fang}
Y.~Jiang, X.~Li, G.~Zhu \emph{et~al.}, ``{Integrated Sensing and Communication
  for Low Altitude Economy: Opportunities and Challenges},'' \emph{IEEE Commun.
  Maga.}, pp. 1--7, 2025.

\bibitem{disadv2022khe}
M.~Khelifi and I.~Butun, ``{Swarm Unmanned Aerial Vehicles (SUAVs): A
  Comprehensive Analysis of Localization, Recent Aspects, and Future Trends},''
  \emph{Journal of Sensors}, vol. 2022, no.~1, p. 8600674, 2022.

\bibitem{tdoa2022sinha}
P.~Sinha and I.~Guvenc, ``{Impact of Antenna Pattern on TOA Based 3D UAV
  Localization Using a Terrestrial Sensor Network},'' \emph{IEEE Trans. on Veh.
  Tech.}, vol.~71, no.~7, pp. 7703--7718, 2022.

\bibitem{sensor2019zhao}
Y.~Zhao, Z.~Li, B.~Hao \emph{et~al.}, ``{Sensor Selection for TDOA-Based
  Localization in Wireless Sensor Networks With Non-Line-of-Sight Condition},''
  \emph{IEEE Trans. on Veh. Tech.}, vol.~68, no.~10, pp. 9935--9950, 2019.

\bibitem{sensor2022zhao}
Y.~Zhao, N.~Cheng, Z.~Li \emph{et~al.}, ``An efficient sensor selection
  algorithm for tdoa localization with estimated source position,'' in
  \emph{2022 IEEE ICC}, 2022, pp. 871--876.

\bibitem{sensor2021dai}
Z.~Dai, G.~Wang, and H.~Chen, ``Sensor selection for tdoa-based source
  localization using angle and range information,'' \emph{IEEE Trans. on Aero.
  and Ele. Sys.}, vol.~57, no.~4, pp. 2597--2604, 2021.

\bibitem{baseTdoa2015sang}
S.~Kim and J.-W. Chong, ``{An Efficient TDOA-Based Localization Algorithm
  without Synchronization between Base Stations},'' \emph{Inter. Journal of
  Dist. Sen. Net.}, vol.~11, no.~9, p. 832351, 2015.

\bibitem{baseTdoa2022fan}
S.~Fan, W.~Ni, H.~Tian \emph{et~al.}, ``Carrier phase-based synchronization and
  high-accuracy positioning in 5g new radio cellular networks,'' \emph{IEEE
  Trans. on Commun.}, vol.~70, no.~1, pp. 564--577, 2022.

\bibitem{baseTdoa2024Motie}
S.~Motie, H.~Zayyani, M.~Salman \emph{et~al.}, ``Self uav localization using
  multiple base stations based on tdoa measurements,'' \emph{IEEE Wire. Commun.
  Letters}, vol.~13, no.~9, pp. 2432--2436, 2024.

\bibitem{3gpp-tr36.777}
{3GPP}, ``Study on enhanced lte support for aerial vehicles,'' 3GPP, Tech. Rep.
  TR 36.777, Dec. 2017, release 15.

\bibitem{bmvd2021drones}
\BIBentryALTinterwordspacing
G.~F.~M. for Digital and Transport. (2021, November) Eu rules for drones.
  [Online]. Available:
  \url{https://dipul.de/homepage/en/aktuelle-meldungen/artikel-3/}
\BIBentrySTDinterwordspacing

\bibitem{3gpp.38.901}
{3GPP}, ``{Study on channel model for frequencies from 0.5 to 100 GHz},'' 3GPP,
  Tech. Rep. TR 38.901, 2020, release 16.

\bibitem{ZYGMsecureUAV2018}
Y.~Zhu, G.~Zheng, and M.~Fitch, ``{Secrecy Rate Analysis of {UAV}-Enabled
  mmWave Networks Using Matérn Hardcore Point Processes},'' \emph{IEEE JSAC},
  vol.~36, no.~7, pp. 1397--1409, 2018.

\bibitem{Venus2020ToA}
A.~Venus, E.~Leitinger, S.~Tertinek \emph{et~al.}, ``Reliability and
  threshold-region performance of toa estimators in dense multipath channels,''
  in \emph{2020 IEEE ICC Workshops}, 2020, pp. 1--7.

\bibitem{Hechen2024ToA}
S.~Hechenberger, S.~Tertinek, and H.~Arthaber, ``Performance bounds of uwb toa
  estimation in presence of wi-fi 6e wideband interference,'' in \emph{2024
  14th IPIN}, 2024, pp. 1--6.

\bibitem{WGKlaus2016}
K.~Witrisal, E.~Leitinger, S.~Hinteregger \emph{et~al.}, ``Bandwidth scaling
  and diversity gain for ranging and positioning in dense multipath channels,''
  \emph{IEEE Wire. Commun. Letters}, vol.~5, no.~4, pp. 396--399, 2016.

\bibitem{GD2025fang}
Z.~Fang, B.~Han, and H.~D. Schotten, ``{Trustworthy UAV Cooperative
  Localization: Information Analysis of Performance and Security},'' \emph{IEEE
  Trans. on Veh. Tech.}, pp. 1--16, 2025.

\end{thebibliography}

\end{document}